\documentclass[twocolumn,secnumarabic,superscriptaddress,amssymb, nobibnotes, aps, prd]{revtex4-2}
\usepackage[hidelinks]{hyperref}

\usepackage{graphicx}

\begin{document}

\title{Photoemission study of Iodoalkanes}%

\author{Maximilian Forster}%
\author{Maximilian Pollanka}%
\author{Sven-Joachim Paul}%
\author{Pascal Freisinger}%
\author{Christian Schröder}%
\affiliation{Chair for Laser- and X-Ray Physics E11, Department of Physics, TUM School of Natural Sciences, Technische Universität München, James-Franck-Straße 1, 85748 Garching, Germany}
\author{Robert Richter}
\author{Michele Alagia}
\affiliation{CNR - Istituto Officina dei Materiali (IOM), Laboratorio TASC, Area Science Park Basovizza, Trieste, Italy.}
\author{Alessio Bruno}
\author{Stefano Stranges}
\affiliation{Department of Chemistry and Technologies of Drugs, Sapienza University, Rome I-00183, Italy.}
\affiliation{CNR - Istituto Officina dei Materiali (IOM), Laboratorio TASC, Area Science Park Basovizza, Trieste, Italy.}
\author{Reinhard Kienberger}
\affiliation{Chair for Laser- and X-Ray Physics E11, Department of Physics, TUM School of Natural Sciences, Technische Universität München, James-Franck-Straße 1, 85748 Garching, Germany}
\email[Contact author: ]{maxi.forster@tum.de}

\date{\today}%
\begin{abstract}
Core level atomic photoionization is influenced by the molecular environment of the emitting atom. To probe this influence we used extreme ultraviolet (XUV) radiation to measure the I$\left(4d\to\varepsilon f \right)$ photoemission of four iodine substituted alkanes in the range from $60\,\mathrm{eV}$ to $160\,\mathrm{eV}$ photon energy. Using two detection angles relative to light polarization allowed us to extract and compare the asymmetry parameter of the investigated compounds. These results are in agreement with theory by \citet{Pi.2018} and previous experiments by \citet{Olney.1998} for iodomethane. 2-iodopropane behaves nearly identical to iodomethane, while for iodoethane and 1-iodopropane we find a distinct and consistent, energy dependant difference most prominent at around 130 eV photon energy. This further proves the influence of the chemical environment on core level photoionization. Due to the sensitivity on the exact positions of the carbon atoms and the energy dependence we hypothesise this difference is caused by intramolecular scattering of the outgoing photoelectron wave packet, causing changes in the angular distribution.
\end{abstract}
\maketitle

\section{Introduction}

Recently, photoemission time delay measurements on the attosecond timescale have been advanced from solids and atoms in the gas phase to molecules. With the two typical methods to assess these delays being RABBIT \cite{Loriot.2024, Makos.2025} and the attosecond streak camera \cite{Ossiander.2018,Biswas.2020, Schroder.2025}. 
\newline
Iodine has repeatedly played a central role in these measurements due to its giant dipole resonance in the I$\left(4d\to\varepsilon f \right)$ channel greatly increasing the photoelectron yield in the energy range from $80\,\mathrm{eV}$ and $120\,\mathrm{eV}$, which is especially relevant in attosecond streaking experiments. Atomic iodine being difficult to obtain, many of these experiments have been performed on small organic molecules containing iodine. The measured time delay, in the range of a few attoseconds, can be related to a scattering phase shift the electron experiences \cite{Wigner.1955, Huppert.2016}. As such, it is expected that the measured delays relate to structural properties of the molecule under study, also investigated theoretically \cite{Yao.2026}. What these experiments typically cannot access readily though is the quantitative information on the photoionization cross-sections and associated asymmetry parameters. With the existence of a link between photoionization dynamics, cross-section and photoelectron angular distribution being well-established (e.g. \cite{Kheifets.2023,Patanen.2013, Dill.1976}), it is obvious that accurate asymmetry parameter- and cross-section data are vital to the interpretation of photoemission timing experiments. Here, we study iodomethane, iodoethane as well as 1-iodopropane and 2-iodopropane in order to supplement the recent experimental studies of \citet{Schroder.2025} and \citet{Biswas.2020}.

\section{Methods}
\subsection{Experimental setup}
\begin{figure}
	\includegraphics[width=0.5\textwidth]{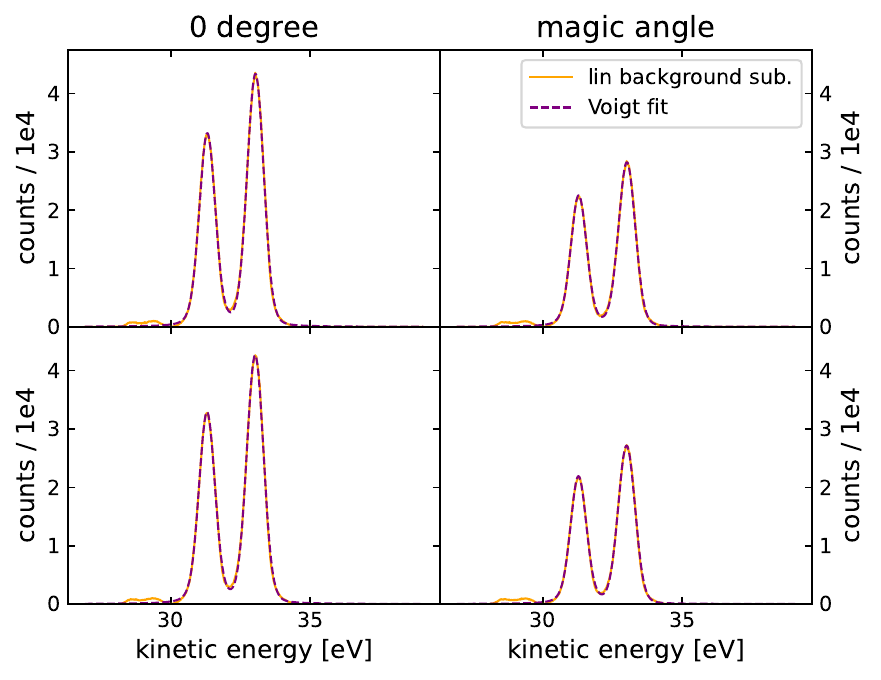}
	\caption{Exemplary data point, four scans taken at two angles with Iodomethane at 90 eV photon energy with Voigt fits. Measurements like this have been performed for every kinetic energy and molecule shown.}
	\label{expamplespec}
\end{figure}

The experiments were conducted at the Gas Phase Photoemission (GAPH) beamline \cite{Prince.1998,Alagia.2003} of the Elettra synchrotron, using the ARPES-TPS end station \cite{Schio.2020}. XUV radiation in the range of $60-150\,\mathrm{eV}$, similarly to the attosecond streaking experiments of \cite{Schroder.2025} which were performed at $90\,\mathrm{eV}$, $105\,\mathrm{eV}$ and $118\,\mathrm{eV}$, has been utilized. The photon energy range was rastered with 5 and 10 eV steps, depending on the expected dynamic at the energy region.
\newline
The experimental chamber features an effusive beam that can transport probes into the vacuum system. The interaction zone is differently pumped multiple times and enables pressures of the order $10^{-7}$ mbar, with much higher pressure in the interaction zone. The samples were commercially bought and trapped gas was removed with the Schlenk technique, namely repeated freeze-thaw cycles under vacuum. They were verified by comparing the valence spectra to literature data \cite{Boschi.1974}. The photoelectrons were detected with the hemispherical analyser using mostly a stepsize of 0.02 eV with an integration time of 1 second. 
\newline
The whole experimental chamber, including the detector, can rotate around the z axis, allowing a quick switch between detection angles, in our case 0$^\circ$ and the magic angle 54.7$^\circ$ relative to light polarization. This technique avoids systematic errors that can arise when using two separate detectors. To account for fluctuations, two scans were taken per angle in an alternate way, this also gives a measure of the statistical error. Photon flux, pressure and integration time were also logged and utilized to normalize the data. In order to reduce the effect of long term drifts, photon energies were measured in a random sequence. An example for one datapoint is shown in figure \ref{expamplespec}. The well studied iodomethane (other name methyl iodide, CH3I) served as a benchmark for our experimental procedure by comparison with published experimental data (e.g. \cite{Olney.1998, Lindle.1984}).
\subsection{Evaluation}
The excellent signal to noise ratio justified a simple linear background subtraction. The spin orbit splitting of the I4$d$ sublevel can be easily resolved over the investigated spectral range. Further splitting, such as by Ligand Field effects (cf. \cite{Cutler.1992}) could not be observed. From the fitted Gaussian $\sigma$ and Lorentzian $\gamma$ we calculated the Gaussian part $\eta$ following the approach by \citet{Thompson.1987} which estimates the Gaussian and Lorentzian parts of a Voigt function. This shows our features are mostly Gaussian-dominated with an average standard deviation of $\sigma = 0.3 \mathrm{eV}$. This corresponds to a resolution limit of about $0.7 \mathrm{eV}$ and is therefore not sufficient to catch fine splitting features. So, for the purposes of our analysis, each peak was fitted with a singular Voigt function, the retrieved amplitude then used as measure of the photoemission cross section. The differential cross section for randomly oriented molecules in dipole approximation is given by \cite{Yang.1948}:
\begin{equation}
	\frac{d\sigma}{\Omega} = \frac{\sigma}{4\pi}\left[1 + \beta P_2(cos\theta)\right],
\end{equation}
with $\theta $ being the angle between the light polarization and the direction of the emitted electron, $P_2$ being the second order Legendre polynomial and the asymmetry parameter $\beta $, which we use to infer $\beta$ from consecutive measurements of the photoemission cross-section at $0\deg$ and at the magic angle.

\section{Results}
\subsection{Photoionization Cross Section of the I4$d$ subshell}
\begin{figure}[t]
	
	\includegraphics[width=0.5\textwidth]{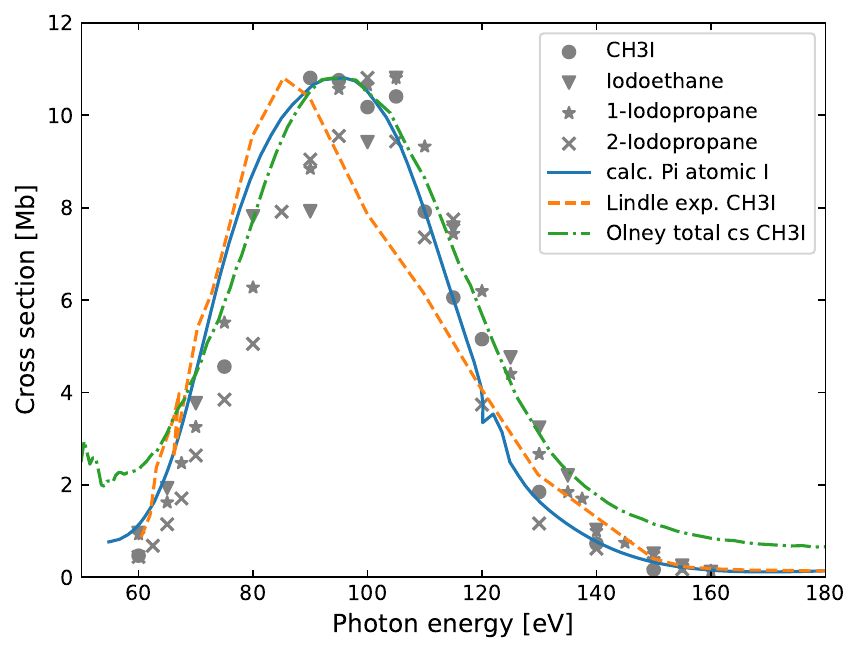}
	\caption{Cross section of the four measured molecules compared to theoretical values \cite{Pi.2018} and previous CH3I measurements \cite{Lindle.1984} \cite{Olney.1998}. Normalized to the absorption measurement of O'Sullivan \cite{OSullivan.1982}. Error bars are barely visible and therefore not shown here.}
	\label{crosssection}
\end{figure}

Normalizing the count rate for each molecule to previous absorption experiments by \citet{OSullivan.1982} yields access to the relative, energy dependant photoionization cross section of the I4$d$ shell. Using the average peak height of the spin orbit split structure measured at the magic angle, shown in figure \ref{crosssection}, where the observed yield is independent from the asymmetry parameter $\beta$. For iodomethane we find reasonable agreement with the experiment of \citet{Lindle.1984} and excellent agreement with that of \citet{Olney.1998}. We find slight variations between the different molecules, with no clear trend. This could well be due to the susceptibility to long term drifts, therefore we don't discuss it further.
 
\begin{figure}
	
	\includegraphics[width=0.5\textwidth]{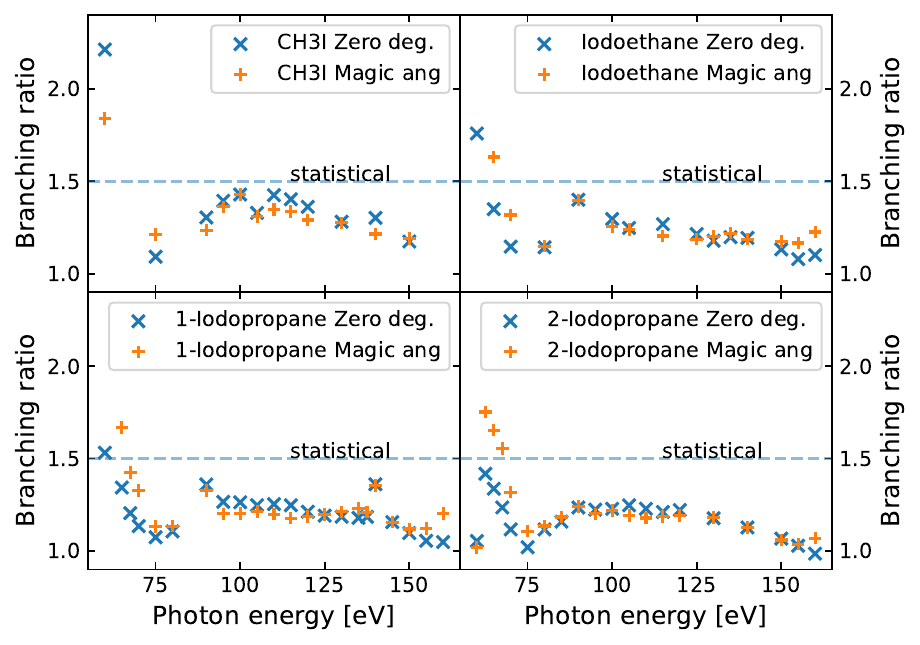}
	\caption{Branching ratio per molecule with the two detection angles plotted separately. While towards higher energies the value is below the statistical factor no obvious systematic difference between the molecules can be observed.}
	\label{branching}
\end{figure}

As mentioned before the used resolution allowed us to resolve the spin orbit split structure for all photon energies, giving as access to the branching ratio $4d_{5/2}$ to $4d_{3/2}$. In figure \ref{branching} the spin orbit branching ratio is shown for each molecule both at $0\deg$ and the magic angle. Here no clear difference between the molecules can be observed. Close to threshold we observe the well-known deviation from the statistical factor $(\ell + 1) / \ell$, at higher energies it is consistently below the statistical factor, contrary to previous data \cite{Lindle.1984}.
\subsection{Chemical shift}
A closer look at the binding energy $E_B$ reveals no significant chemical shift for the investigated molecules (see table \ref{be}). The differences are well within the margin of error, showing that variation of the extended functional group does not significantly influence the iodine core state.
\begin{table}

\caption{Binding energy of the investigated molecules in eV with statistical error.}
\label{be}
\begin{center}
\begin{tabular}{l | c | c}
	Molecule & $E_b^{3/2}$ [eV] & $E_b^{5/2}$ [eV] \\
	\hline
	Iodomethane   & $ 58.76 \pm 0.25$ & $ 57.05 \pm 0.07 $\\
	Iodoethane    & $ 58.59 \pm 0.26$ & $ 56.97 \pm 0.21 $\\
	1-Iodopropane & $ 58.53 \pm 0.25$ & $ 56.88 \pm 0.22 $\\
	2-Iodopropane & $ 58.27 \pm 0.20$ & $ 56.64 \pm 0.19 $\\
	
\end{tabular}
\end{center}

\end{table}
\subsection{Asymmetry parameter}
\begin{figure}
	\includegraphics[width=0.5\textwidth]{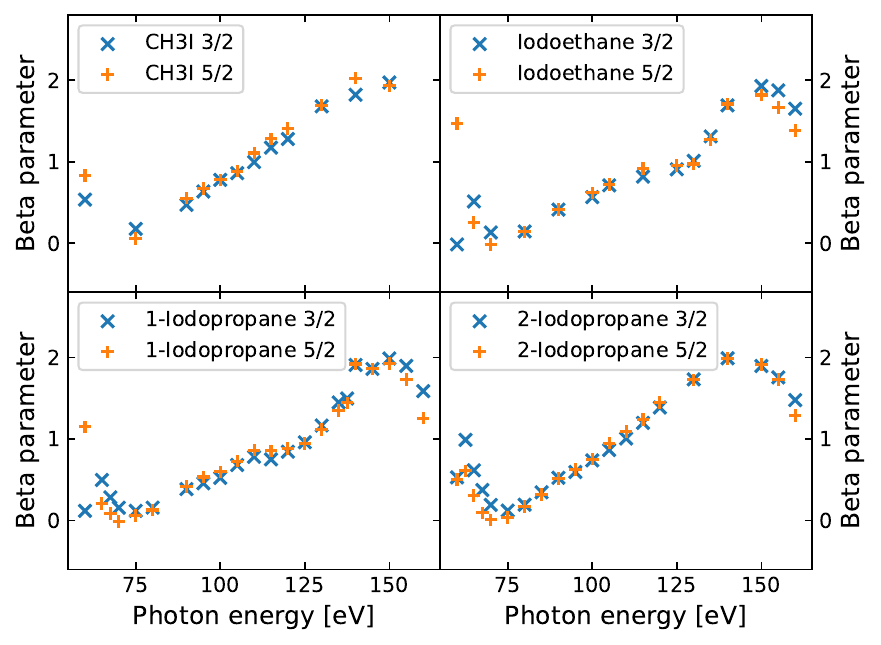}
	
	\caption{Beta parameter spin orbit resolved. Apart from the very low and high photon energies the two states yield nearly identical values. }
	\label{beta_sep}
\end{figure}

The asymmetry parameter of iodomethane follows previously obtained experimental and theoretical results rather well, reproducing the general form, see figure \ref{beta_sep}. For most energies, the beta parameter for both states $4d_{5/2}$ and $4d_{3/2}$ is practically identical, especially in the targeted range from $80 \,\mathrm{eV}$ to $140 \, \mathrm{eV}$. In iodomethane we do not observe the disruption from a linear increase around 120 eV as has been found in \cite{Lindle.1984}. This has been attributed to the I$4p$ state bound at around $129 \mathrm{eV}$ \cite{Forbes.2020} and enabling autoionization of Rydberg states. 2-iodopropane behaves practically identical to iodomethane. These two molecules show a remarkably smooth linear increase from $80 \,\mathrm{eV}$ to $140 \, \mathrm{eV}$. Most notably, iodoethane and 1-iodopropane deviate from this linear increase. Starting at around $100 \,\mathrm{eV}$ their incline is less steep and at 110 eV they have a significant dent. At around 120 eV they have a large difference in the $\beta$ value with respect to the other two molecules, above 0.5. Towards higher energies the curves rejoin. The four molecules are split into two distinct groups that behave nearly identical.

\begin{figure}
	
	\includegraphics[width=0.5\textwidth]{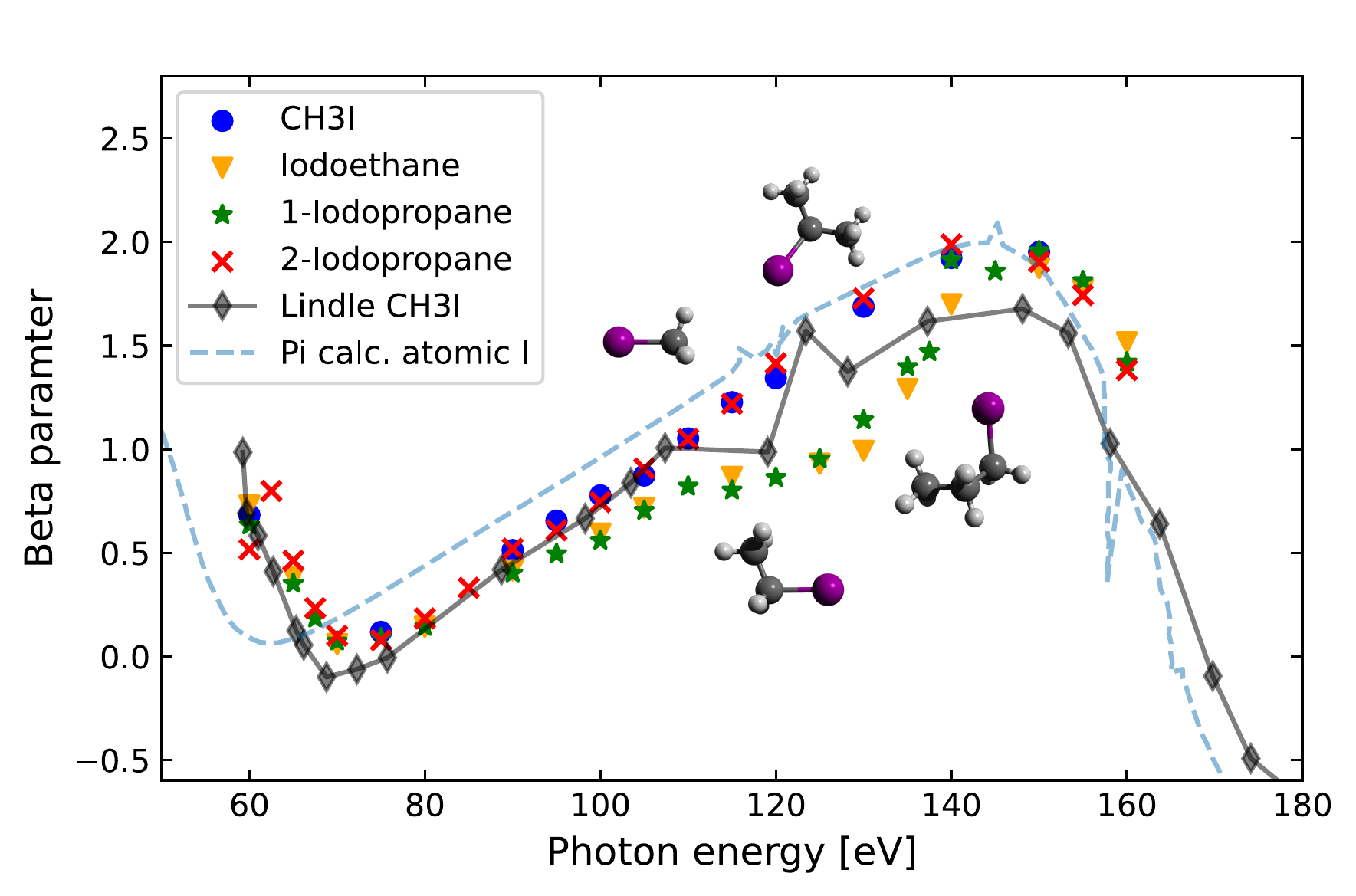}
	\caption{Beta parameter of the 4 investigated molecules compared to older data \cite{Lindle.1984} and theory \cite{Pi.2018}. The distinction between the two groups of molecules is clearly visible in the energy range between 100 and 140 eV.	}
	\label{fig1}
\end{figure}

\section{Discussion}
The behaviour of the asymmetry parameter, analogously to the photoemission time delay, evades an intuitive explanation. Not only the size of the molecule but also the position of the emitting atom within the molecule is evidently important. This becomes especially clear when comparing 1-iodopropane to 2-iodopropane. This, together with the energy dependence of the observed effect, supports the idea that the outgoing photoelectron scatters elastically when leaving the molecule, changing its direction and phase and the process being very sensitive to the exact position of the scattering centers. An argument can be made that in 2-iodopropane the effects of both carbon atoms cancel each other out due to the equivalent distance from both atoms to the iodine atom, thereby yielding the same result as iodomethane. 
\newline
Noteworthy is also the effects that various theoretical approaches have observed in the energy region around $120 \,\mathrm{eV}$ such as for iodomethane in \cite{Pi.2018}. \citet{Kheifets.2026} have investigated the four molecules of this work and found differences in the shape of the asymmetry parameter depending on the molecule in this energy region.
\newline
While the molecules have different symmetries the asymmetry parameter does not change along the symmetry groups. Iodomethane has the highest symmetry, belonging to the C\textsubscript{3v} symmetry point group, iodoethane and 2-iodopropane have only one symmetry element (one plane of symmetry) and both belong to the C\textsubscript{s} point group,  and 1-iodopropane is an asymmetric molecule with no symmetry elements, therefore belonging to the lowest symmetry C\textsubscript{1} point group, being in this case the identity operator the only symmetry operator of the molecule. However the point can be made that from the point of view of the iodine atom in 2-iodopropane two carbon atoms have equal distance, which is not given in iodoethane. For the I4$d$ photoemission this most likely plays a more important role than the symmetry of the whole molecule itself. 
\section{Conclusion and outlook}
Over 400 individual spectra have been taken in the XUV range to systematically investigate the asymmetry parameter across four different iodine containing molecules to gain insight in the how the ligands changes the I$4d$ photoemission. We found a large, distinctive difference between two of the molecules respectively. The asymmetry parameter of iodoethane and 1-iodopropane deviates between $110$ and $140 \mathrm{eV}$  from the other two molecules and from the theoretical data obtained for atomic iodine. This strongly suggests that the observed effect is due to molecular effects, such as intramolecular scattering, that depend on the exact positions of the scattering centers and the energy of the outgoing photoelectron wave packet.
\newline
Due to time constraints iodobuthane (1- and 2-) has not been measured in this beamtime, which could provide more insight. A higher energy resolution and tighter sampling might give more insights into the exact dynamics appearing in the photon energy region of $110 \,\mathrm{eV}- 140\, \mathrm{eV}$.
\section{Acknowledgement}
The experiments were conducted at the Elettra synchrotron Trieste. We thank the late Robert Richter for assistance in planning and executing the measurement campaign and the  inspiring discussions.

\bibliography{Photoemission_study_of_Iodoalkanes.bib}
\end{document}